\documentclass{article}
\usepackage{amssymb}

\usepackage{amsfonts}
\usepackage{epsfig}

\newcommand{\be}{\begin{equation}}
\newcommand{\ee}{\end{equation}}

\begin{document}

\title{Electromagnetically induced switching of ferroelectric thin films}
\author{J.-G.~Caputo$^{1~2~*)}$, E. V.~Kazantseva$^3$ and A.I.~Maimistov$%
^4$}
\maketitle

\begin{center}
{\normalsize \textit{1) Laboratoire de Math\'ematiques, INSA de Rouen,}}\\[%
0pt]
{\normalsize \textit{B.P. 8, 76131 Mont-Saint-Aignan cedex, France.}} \\[0pt]
{\normalsize \textit{\phantom{1)} E-mail: caputo@insa-rouen.fr}} \\[0pt]
{\normalsize \textit{2) Laboratoire de Physique th\'eorique et modelisation,}%
}\\[0pt]
{\normalsize \textit{Universit\'e de Cergy-Pontoise and C.N.R.S.}}\\[0pt]
{\normalsize \textit{3)Laboratoire de Physique de l'Universit\'{e} de
Bourgogne,}} \\[0pt]
{\normalsize \textit{Av.Alain Savary, 9, 21078 Dijon, France}} \\[0pt]
{\normalsize \textit{\phantom{1)} E-mail: murkamars@hotmail.com }}\\[0pt]
{\normalsize \textit{4) Department of Solid State Physics,}} \\[0pt]
{\normalsize \textit{Moscow Engineering Physics Institute,}}\\[0pt]
{\normalsize \textit{Kashirskoe sh. 31, Moscow, 115409 Russia}}\\[0pt]
{\normalsize \textit{\phantom{1)} E-mail: amaimistov@hotmail.com}}
\end{center}



\begin{abstract}

We analyze the interaction of an electromagnetic
spike (one cycle) with a thin layer of ferroelectric medium with two equilibrium states. 
The model is the set of Maxwell equations 
coupled to the undamped Landau-Khalatnikov equation, where we do
not assume slowly varying envelopes. From linear scattering theory, we show 
that low amplitude pulses can be completely reflected by the medium.
Large amplitude pulses can switch the ferroelectric. Using numerical 
simulations and analysis, we study 
this switching for long and short pulses, estimate the switching times and
provide useful information for experiments.

\end{abstract}

\section{Introduction}

Recently there has been a revival of interest in the study of 
phenomena occurring during the propagation of electromagnetic 
waves in different media having a long-range order. The propagation of 
electromagnetic pulses (in particular, of optical range) as stable solitary
waves, sometimes called solitons, occupies a special place in such
investigations. 
As an example, we can point out to studies of 
electromagnetic solitons propagating in ferro(antiferro)magnetic media \cite%
{R1,R2,R3}. Based on the Landau-Lifshitz equations for magnetization and the
Maxwell equations for the electromagnetic field, solitary waves (supersonic, as
distinct from magnetic solitons) were found. These can be approximately
described by the modified Korteweg-de Vries (mKdV) equation. It was shown
that they are stable against structural perturbations of the mKdV equation
caused by energy dissipation. Using the multiscale
perturbation theory, electromagnetic solitons were considered in
antiferromagnetic \cite{R4} and anisotropic ferromagnetic media \cite{R5}.
It was shown in \cite{R6} that the modulation of an electromagnetic wave in
a ferromagnet in an external field can be described by the nonlinear Schr%
\"{o}dinger equation. This opens the way to study the modulational
instability and the formation of electromagnetic solitons in magneto-ordered
media.

Dielectrics having a permanent polarization in the absence of an external
electric field, called pyroelectrics, are, in many respects,
similar to magneto-ordered media. If the phase transition between 
the pyroelectric and the nonpyroelectric states is of second-order, such
pyroelectrics are called \emph{ferroelectrics}. The Landau phenomenological
theory of phase transitions is based on the assumption of an
order parameter that is zero in one phase and is nonzero in the other
phase. A uniaxial ferroelectric provides a simple example of this theory of
phase transitions. Due to the nonlinearity of the free energy of the
ferroelectric relative to the order parameter, there can exist nonlinear
waves of spontaneous polarization both in the form of solitons \cite{R7} -%
\cite{R10} and in the form of domain walls \cite{R11}. Thin films
of ferroelectrics \cite{R12, R13} and liquid crystals having ferroelectric
properties \cite{R14, R15,Ntogari} are of interest for practical use. This
is because such media, having a permanent dipole
moment, can be used to generate efficiently optical harmonics
\cite{R14}. Ferroelectrics can also be used to create memory 
devices and optically controlled switches 
\cite{R13, Ntogari, Pic, Isak}. Many recent results are collected in
the review on two dimensional ferroelectrics and thin polymer 
ferroelectric films represents in \cite{Blin}.

Recently \cite{R20} we
investigated the traveling wave solutions of the Maxwell-Duffing
homogeneous system describing the
interaction of short electromagnetic pulses with a bulk ferroelectric.
The description of the fast switching and ultra-short pulse propagation
requires that the duration of these 
pulses be shorter than the relaxation time of the nonequilibrium 
polarization, which, for ferroelectrics, is equal to 
several nanoseconds \cite{R13}. 
In this study, we consider the interaction of an extremely short light
pulse with a thin film of dielectric medium having a spontaneous polarization (a
ferroelectric). Using the \emph{phenomenological
Landau-Khalatnikov model} \cite{R16,R18, R19,Lu} describing a uniaxial
ferroelectric and the Maxwell equation for an
electromagnetic wave, we consider the reflections and refractions 
of a short electromagnetic spike (i.e., a pulse without a carrier wave) through 
a ferroelectric thin film. We assumed that the film width is less than the
spatial size of the spike, but is greater then the critical length 
$L_{c}$ below which no ferroelectricity exists \cite{Pic, Lu, Blin}. 
According to theoretical estimations and 
\cite{Lu, R13} $L_{c}=0.5 nm$ for $BaTiO_3$, $L_{c}=20 nm$ for $PbTiO_3$. 
For experimental films of two or five monolayers of polyvinylidene 
fluoride (PVDF) and copolymer P(VDF-TrFE) 
$L_{c}\approx 7 nm $ \cite{Blin}.

The article is organized in the following way. Section 2 presents the 
model together with a solution of the Maxwell equations for a localized
ferroelectric medium. We apply these results to a thin film in section
3. Section 4 is devoted to the switching caused by large amplitude 
electromagnetic pulses. We conclude in section 5.

\section{Phenomenology of ferroelectricity}

A phenomenological description of ferroelectricity due to Landau and
Khalatnikov \cite{R16,R18,Kittel} gives the following Lagrangian density for
the interaction of an electromagnetic field and a dielectric medium
\begin{equation}
L={\frac{1}{8\pi c^2}}(\frac{\partial \mathbf{A}}{\partial t})^{2}
-{\frac{1}{8\pi }}(\mathrm{rot}\mathbf{A})^{2}
+I(\mathbf{{x})} \left[ \frac{1}{2g} ( 
\frac{\partial \mathbf{P}}{\partial t})^{2}-\frac{1}{g}\Phi (\mathbf{P})
-{\frac{1}{c}}\mathbf{P}\frac{\partial \mathbf{A}}{\partial t}  
\right] 
\label{lag_ap}
\end{equation}%
where $\mathbf{A}$ is the vector potential, $\mathbf{P}$ is the polarization
of the medium, $c$ the speed of light, $g$ is coupling constant, $\Phi (%
\mathbf{P})$ is the thermodynamic potential, the last term describes the
coupling between $\mathbf{A}$ and $\mathbf{P}$. The ferroelectric medium can
exist only in some region and therefore we have introduced in (\ref{lag_ap})
$I(\mathbf{{x})}$, the characteristic function of the medium, i.e., $I(%
\mathbf{x})=1$ inside the medium and $I(\mathbf{x})=0$ outside. The
Lagrangian approach guarantees that we take into account the correct
couplings.

\subsection{Homogeneous case}

We first consider the homogeneous situation 
i.e. we assume $I(\mathbf{x})\equiv 1$. The variation of the action
functional yields the equations for $\mathbf{A}$ and $\mathbf{P}$%
\[
\mathrm{rot}\mathrm{rot}\mathbf{A}+{\frac{1}{c^{2}}}\frac{\partial ^{2}%
\mathbf{A}}{\partial t^{2}}=-{\frac{4\pi }{c^{2}}}\frac{\partial \mathbf{P}}{%
\partial t},
\]%
\[
{\frac{\partial ^{2}\mathbf{P}}{\partial t^{2}}}+{\frac{{\delta }\Phi (%
\mathbf{P})}{{\delta }\mathbf{P}}}=-{\frac{g}{c^{2}}}\frac{\partial \mathbf{A%
}}{\partial t},
\]
which when written in terms of the electric field $\mathbf{E}=-(1/c)\partial
\mathbf{A}/\partial t$ \ are
\begin{equation}
\mathrm{rot}\mathrm{rot}\mathbf{E}+{\frac{1}{c^{2}}}\frac{\partial ^{2}%
\mathbf{E}}{\partial t^{2}}=-{\frac{4\pi }{c^{2}}}\frac{\partial ^{2}\mathbf{%
P}}{\partial t^{2}},  \label{eqe}
\end{equation}%
\begin{equation}
{\frac{\partial ^{2}\mathbf{P}}{\partial t^{2}}}+{\frac{{\delta }\Phi (%
\mathbf{P})}{{\delta }\mathbf{P}}}=g\mathbf{E}.  \label{eqp}
\end{equation}%
We can introduce a phenomenological relaxation of the medium polarization by
adding to the right hand side of (\ref{eqp}) a linear damping term and
obtain the well-known Landau-Khalatnikov equation.

To simplify the problem we will consider an electric field polarized along
the propagation variable $x$ so that $\mathbf{E}=E\mathbf{e_{x}}$ and a
polarization along $x$ $\mathbf{P}=P\mathbf{e_{x}}$. Following Landau the
potential can be chosen as
\begin{equation}
\Phi (P)=\Phi _{0}+{\frac{1}{2}}\alpha P^{2}+{\frac{1}{4}}\beta P^{4}+{\frac{%
1}{2}}D(\frac{\partial P}{\partial x})^{2}.  \label{phiofp}
\end{equation}%
In the theory of Landau $\alpha =\alpha _{0}(T-T_{c})$ depends on the
temperature and not $\beta $. If $\alpha >0$ the potential is minimum for $%
P=0$ and this corresponds to a (disordered) paraelectric phase. On the
contrary if $\alpha <0$ and $\beta >0$ there are two minima located at $%
P=\pm \sqrt{-\alpha /\beta }$ corresponding to two opposite orientations
of the polarization, this is the ferroelectric phase. Then the
Euler-Lagrange equations for $E,P$ are
\begin{eqnarray}
{\frac{1}{c^{2}}}\frac{\partial ^{2}E}{\partial t^{2}}-\frac{\partial ^{2}E}{%
\partial x^{2}} &=&-{\frac{4\pi }{c^{2}}}\frac{\partial ^{2}P}{\partial t^{2}}
\label{maxduf} \\
\frac{\partial ^{2}P}{\partial t^{2}}-D\frac{\partial ^{2}P}{\partial x^{2}}%
+\alpha P+\beta P^{3} &=&gE.  \nonumber
\end{eqnarray}%
Traveling wave solutions for the system (\ref{maxduf}) were found in \cite%
{R20}, they include a one-parameter family of solitons, a one-parameter
family of kinks (domain wall) solutions and a one-parameter family of
periodic (cnoidal) waves. In \cite{R20} we found numerically that the bright solitons on a
zero and nonzero polarization background are stable while the dark solitons
are stable only on a zero background.

\subsection{Inhomogeneous case}

We will now normalize the fields and variables as
\be \label{norma}
t' = t/\sqrt{|\alpha|},~~x'= x\sqrt{|\alpha|}/c, ~~
\mathbf{A} = 2 c \sqrt{\pi |\alpha| \over g \beta} \mathbf{a},~~
\mathbf{P}={|\alpha| \over \beta}\mathbf{q},\ee
where the polarization $P$ is normalized by the saturation value.
The Lagrangian density (\ref{lag_ap}) becomes
\be\label{lagnorm}
L={\frac{1}{2}}(\frac{\partial \mathbf{a}}{\partial t})^{2}
-{\frac{1}{2 }}(\mathrm{rot}\mathbf{a})^{2}
+I(\mathbf{{x})} \left[ \frac{1}{2} (
\frac{\partial \mathbf{q}}{\partial t})^{2}
+ m\frac{\mathbf{q}^2 }{2}
-\frac14\mathbf{q}^4 
-\gamma \mathbf{q}\frac{\partial \mathbf{a}}{\partial t}\right]
\ee
where the primes have been dropped. The parameters are
\be\label{param}
\gamma = 2 \sqrt{\pi g \over |\alpha|}~~,~~m=\alpha/|\alpha|,\ee
where $m$ is the "mass" of the excitations which can be negative for
ferroelectrics and positive for paraelectrics and $\gamma $ is the
polarisability of the medium. 
In the theory of Landau-Khalatnikov, the parameter $g$ is the
susceptibility of the material so $g=1/(4|\alpha|)$ in the ferroelectric 
phase. From (\ref{param}) we have
$$\gamma = \sqrt{\pi \over 2}{1\over |\alpha|}= 
\sqrt{\pi \over 2} {1\over |\alpha_0|}|T_c-T|^{-1}.$$
Consider for example a crystal $BaTiO_3$, then $\alpha_0 = 
6\times 10^{-6}K^{-1},~~~
\beta = 2\times 10^{-15}m^3J^{-1}$ so that 
$\gamma = 2 \times 10^5 |T_c-T|^{-1}$.

The simplified Lagrangian density \cite{mc03} corresponding
to a one-dimensional plane wave:
\begin{equation}
\mathcal{L}=
\frac{a_{t}^{2}}{2}
-\frac{a_{x}^{2}}{2}
+I(x)(\frac{q_{t}^{2}}{2}%
+m\frac{q^{2}}{2}+ \frac{q^{4}}{4}-\gamma qa_{t}),  \label{lagdens}
\end{equation}%
where $a$ is the analog of vector potential. 
The Euler-Lagrange equations are
\begin{eqnarray}
a_{tt}-a_{xx} &=&-\gamma I(x)q_{t}  \label{aq} \\
q_{tt}+mq+ q^{3} &=&\gamma a_{t}
\end{eqnarray}%
The equations for the electric field $e=-a_{t}$ and medium variable can then
be obtained
\begin{eqnarray}
e_{tt}-e_{xx} &=&-\gamma I(x)q_{tt},  \label{eq1} \\
q_{tt}+mq+ q^{3} &=&\gamma e,
\end{eqnarray}%
where the coupling between the fields $e$ and $q$ only occurs in the medium
i.e. on the support of $I(x)$.

\begin{figure}[tbp]
\centerline{\psfig{figure=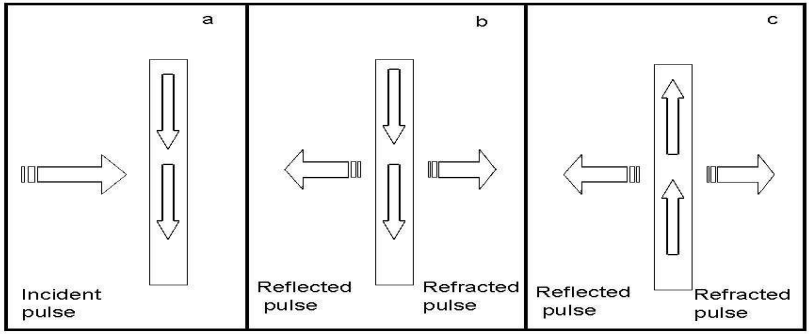,height=6cm,width=12cm,angle=0}}
\caption{Schematic of an incident electromagnetic pulse on a thin ferroelectric
film panel $a$ (left). The $b$ and $c$ panels show respectively the case
of switching and non switching of the polarization.}
\label{fig1}
\end{figure}

We now consider the general scattering formalism assuming a localized
electromagnetic wave impinging on the medium from the left like is shown in
Fig. \ref{fig1}. The general problem can only be treated numerically so we
simplify it and consider two limiting cases, an array of thin films and a
single thin film.

\subsection{Scattering of an electromagnetic wave by a ferroelectric slab}

We assume that the electromagnetic wave is incident from the left $x<0$ on a
medium whose position is given by the indicator function $I(x)$. Then the
wave equation reads
\begin{eqnarray}
e_{tt}-e_{xx} &=&g(x,t)  \label{eg} \\
g(x,t) &=&-\gamma q_{tt}I(x),
\end{eqnarray}%
where the boundary conditions are
\begin{equation}
e(t,x=\pm \infty )=0,~~e_{t}(t,x=\pm \infty )=0,  \label{bc}
\end{equation}%
and the initial conditions
\begin{equation}
e(t=0,x)=e_{0}(x),~~e_{t}(t=0,x)=e_{1}(x)=-\frac{\partial }{\partial x}%
e_{0}(x).  \label{Eqn4}
\end{equation}%
We suppose that the initial pulse $e_{0}(x)$ is located far to the left of
the medium. The equation for the field is linear so to solve we introduce
the Fourier transform of $e$ and similarly for $g$
\begin{equation}
e(x,t)=\int\limits_{-\infty }^{+\infty }\frac{\ dk}{2\pi }\hat{e}(k,t)\exp
(ikx),~~\hat{e}(k,t)=\int\limits_{-\infty }^{+\infty }e(k,t)\exp (-ikx)\ dx,
\end{equation}%
to obtain the initial value problem
\begin{equation}
\hat{e}_{tt}+k^{2}\hat{e}=\hat{g}(k,t),  \label{Eqn5}
\end{equation}%
with $\hat{e}(k,t=0)=\hat{e}_{0}(k),~~~\hat{e}_{t}(k,t=0)=\hat{e}_{1}(k).$

Following the general approach \cite{ts83} we write $\hat{e}=\hat{e}^{0}+%
\hat{e}^{1}$ where $\hat{e}^{0}$ solves the homogeneous equation
\begin{equation}
\hat{e}_{tt}^{0}+k^{2}\hat{e}^{0}=0,  \label{Eqn6}
\end{equation}%
with the initial conditions $\hat{e}^{0}(k,t=0)=\hat{e}_{0}(k),~~\hat{e}%
_{t}^{0}(k,t=0)=\hat{e}_{1}(k).$ and $\hat{e}^{1}$ solves the inhomogeneous
equation
\begin{equation}
\hat{e}_{tt}^{1}+k^{2}\hat{e}^{1}=\hat{g}(k,t),  \label{Eqn7}
\end{equation}%
with $\hat{e}^{1}(k,t=0)=0,~~\hat{e}_{t}^{1}(k,t=0)=0.$

The general solution of (\ref{Eqn6}) is
\[
\hat e^{0} = \frac{1}{2} \left( \hat e_0(k) + \frac{i}{k} \hat e_1(k)
\right) \exp(-ikt) + \frac{1}{2} \left( \hat e_0(k) - \frac{i}{k} \hat
e_1(k) \right) \exp(+ikt),
\]
Using for example a Green's function approach one easily sees that the
solution of the problem (\ref{Eqn7}) is
\[
\hat e^{1} = \int\limits_{0}^{t}\frac{\sin k(t - \tau)}{k} \hat g(k , \tau)
d \tau ,
\]

The solution of the homogeneous problem can be transformed into the
following expression
\[
e^{0}(x,t)=\frac{1}{2}\left[ e_{0}(x-t)+e_{0}(x+t)+\int%
\limits_{x+t}^{x-t}e_{1}(y)dy\right] \equiv e_{0}(x-t)
\]%
the first part of which is D'Alembert's formula. The inhomogeneous solution
can be rewritten as
\[
e^{1}(x,t)=\frac{1}{2}\int\limits_{-\infty }^{+\infty
}\int\limits_{0}^{t}g(y,\tau )\left[ \theta (x-y-t+\tau )-\theta (x-y+t-\tau
)\right] d\tau dy,
\]%
where we use the step-function $\theta (z)=\int\limits_{-\infty }^{z}\delta
(x)dx$. So we have the general solution of the scattering problem under
consideration.

\begin{equation}  \label{Eqn8}
e(x,t) = e_0(x-t) + \frac{1}{2} \int\limits_{-\infty }^{+\infty } \int
\limits_{0}^{t} g(y, \tau) \left[ \theta(x-y-t+ \tau) - \theta(x-y+t - \tau) %
\right] d\tau dy
\end{equation}

As an example we give the result for an array of thin films, useful for
applications and theory. The indicator function is
\begin{equation}  \label{Indicator}
I(x) = \sum \limits_{i=1}^{N} \delta (x-x_{i}),
\end{equation}
and the final result is
\begin{equation}  \label{eatf}
e(x,t) = e_{0}(x-t) + \frac{\gamma}{2}\sum
\limits_{i=1}^{N}\int\limits_{0}^{t} q_{t}(x_{i}, \tau) \left[ \delta(\tau
-t+x-x_{i}) + \delta(\tau -t-x+x_{i}) \right] d\tau,
\end{equation}
where the evolution of the state of every film at point $x = x_{i}$ is
defined by the equation
\[
q_{tt}(x_{i},t) +m q(x_{i},t) + q(x_{i},t)^{3} = \gamma e (x = x_{i},
t).
\]
This problem is still complicated to analyze because of the delays and
advances in equation (\ref{eatf}). We therefore simplify drastically the
situation by considering a single thin film only.

\section{A single thin film}

There is the simplest not trivial case when the nonlinear medium is
represented by a single thin film of anharmonical oscillators. Here we can
calculate the integral in expression (\ref{eatf}) exactly. The electric
field is defined by the following expression
\begin{equation}
\begin{array}{lcl}
e(x,t) & = & e_{0}(x-t)+\frac{\gamma }{2}\int\limits_{0}^{t}q_{t}(0,\tau )%
\left[ \delta (\tau -t+x)+\delta (\tau -t-x)\right] d\tau .%
\end{array}
\label{singl1}
\end{equation}%
At the point where the film is placed ($x=0$) we have
\begin{eqnarray*}
e(0,t) &=&e_{0}(-t)+\frac{\gamma }{2}\int\limits_{0}^{t}q_{t}(0,\tau )\left[
\delta (\tau -t)+\delta (\tau -t)\right] d\tau \\
&=&e_{0}(-t)-\frac{\gamma }{2}q_{t}(0,t).
\end{eqnarray*}%
Substituting this expression into the equation of the oscillator results in
the following equation
\begin{equation}
q_{tt}+mq+ q^{3}=\gamma e_{0}(-t)-\frac{\gamma ^{2}}{2}q_{t}.
\label{single2}
\end{equation}

This model represents the evolution of the nonlinear Duffing oscillator with
both damping and forcing. 

\subsection{Linear considerations}

In the homogeneous case $g\equiv 1$ we can compute the dispersion relation
for the system (\ref{aq}) by assuming $a=a_{0}e^{i(kx-\omega
t)},q=q_{0}e^{i(kx-\omega t)}$ and obtain
\[
k^{2}=\frac{\omega \gamma ^{2}}{1-\omega ^{2}}+\omega ^{2}
\]%
The dispersion relation $\omega (k)$ presents the well known polaritonic gap.

When the medium is localized like for the thin film case, the picture
changes and we need to make a scattering experiment to understand the linear
behavior of the device. The equations (\ref{eq1}) become for the
paraelectric case
\begin{eqnarray}
e_{tt}-e_{xx} &=&-\gamma q_{tt}\delta (x),  \label{eqd1} \\
q_{tt}+q &=&\gamma e\ (x=0,t).
\end{eqnarray}%
We compute the reflection and transmission coefficients of an incident
linear wave on such a medium. For that we introduce the harmonic dependence $%
e(x,t)=e^{i\omega t}E(x),~~ q(t)=Qe^{i\omega t}$. Plugging these into the
previous system of equations yields the Schroedinger equation with delta
function potential
\begin{equation}  \label{schro}
E_{xx} + \left[ \omega^2 +{\frac{\gamma\omega^2 }{1-\omega^2}} \delta(x)%
\right] E = 0 .
\end{equation}
For the scattering we assume a wave incident from the left $x<0~~ E =
e^{-ikx} + Re^{ikx}$ and a transmitted wave $x>0~ E = Te^{-ikx}$. The
continuity of $E$ at $x=0$ and the jump of the derivative $%
[E_{x}]_{0^{-}}^{0^{+}}=-\gamma \omega ^{2} E(0)
/(1-\omega ^{2})$
give the transmission and reflection coefficients
\begin{equation}
T=\frac{2 i (1-\omega ^{2})}{2 i (1-\omega ^{2})-\gamma \omega},
\label{trans}
\end{equation}
\begin{equation}
R = T-1 =\frac{\gamma \omega}{2 i (1-\omega ^{2})-\gamma \omega},
\label{ref}
\end{equation}
where we used the dispersion relation $k=\omega$.

Several remarks can be made. First transparency $R=0$ is obtained only for $%
\omega =0$ and total reflection occurs for $\omega^{2}=1$ as expected
\cite{lamb83}. We also have two
bound states corresponding to the poles of $R$ and $T$
\begin{equation}
\omega =-{\frac{i\gamma ^{2}}{4}}\pm \sqrt{1-{\frac{\gamma ^{4}}{16}}},
\label{poles}
\end{equation}%
which are located in the lower half complex plane.

In the ferroelectric case the medium will oscillate around one of the
equilibria ${\bar q} = \pm 1$ so that the linearized equations
are
\begin{eqnarray}
e_{tt}-e_{xx} &=&-\gamma q_{tt}\delta (x),  \label{lferro} \\
q_{tt}+2 q &=&\gamma e\ (x=0,t).
\end{eqnarray}%
The reflection and transmission coefficients become
\begin{equation}
R=T-1= \frac{\gamma \omega}{2 i (2-\omega ^{2})-\gamma \omega}.
\label{reftraf}
\end{equation}
In this case the poles are given by 
\begin{equation}
\omega =-{\frac{i\gamma ^{2}}{4}}\pm \sqrt{2-{\frac{\gamma ^{4}}{16}}},
\label{poles2}
\end{equation}

\begin{figure}[tbp]
\centerline{\psfig{figure=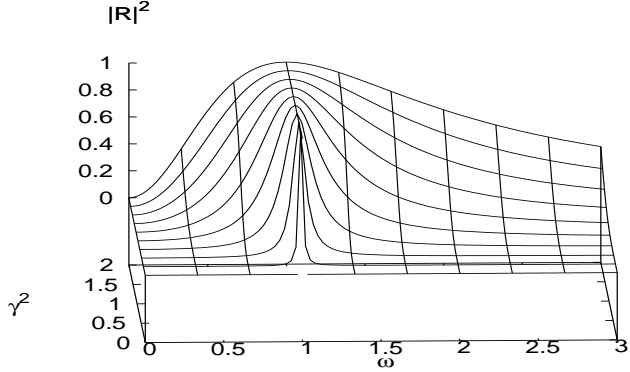,height=6cm,width=12cm,angle=0}}
\caption{Modulus of the reflection coefficient 
$|R (\protect\omega,\protect\gamma^2)|^2$. }
\label{fig1a}
\end{figure}

In Fig. \ref{fig1a} we show the reflection coefficient 
$|R (\omega,\gamma^2)|^2$. It is close to 1 for the resonant frequency $\omega =1$ (resp. $%
\sqrt{2}$) in the paraelectric (resp. ferroelectric) case. The width of the
resonance is the same for both cases and increases when the coupling $\gamma
$ grows. The coefficient $\gamma $ is the coupling between the
electromagnetic field and the medium. It is appears as a damping of the
medium polarization in (\ref{single2}). The mechanism of relaxation is
radiative.

\section{Effect of anharmonicity: switching}

When the amplitude of the incident pulse is large enough the anharmonic term
in (\ref{single2}) needs to be taken into account. If the system is
paraelectric ($m>0$) then it gets kicked out of the equilibrium
state $q=0$ and relaxes back to it. More interesting is the ferroelectric
system ($m<0$) which has two stable equilibria $\bar{q}=\pm 1 $ and one unstable $\bar{q}=0$ so that switching between them is
possible.
We will solve
\begin{equation}
q_{tt}-q+ q^{3}=\gamma e_{0}(-t)-\frac{\gamma ^{2}}{2}q_{t},
\label{ferroq}
\end{equation}%
for an incoming electromagnetic pulse $e_{0}(-t)$ assuming the ferroelectric
medium is in its equilibrium position i.e. with the initial condition $%
q_{t}(t=-\infty )=0,~~q(t=-\infty )=\bar{q}=-1 $.
For a harmonic perturbation, chaotic behavior can occur, see for example 
\cite{gh83}. Here we have a force of
finite duration so we expect transient chaos which will cause irregular
switching.

From the experimental point of view it easier to work with pulses whose
profiles are gaussian or plateau-like form. We chose the plateau-like form
\begin{equation}
e_{0}(-t)=A_{0}\left( \tanh \frac{-t-t_{1}}{t_{f}}-\tanh \frac{-t-t_{2}}{%
t_{f}}\right) .  \label{inpulse}
\end{equation}%
The initial polarization of the ferroelectric medium is defined by the
parameter $q(t=-\infty )=\bar{q}$.

\subsection{Short electromagnetic pulse}

\begin{figure}[tbp]
\centerline{\psfig{figure=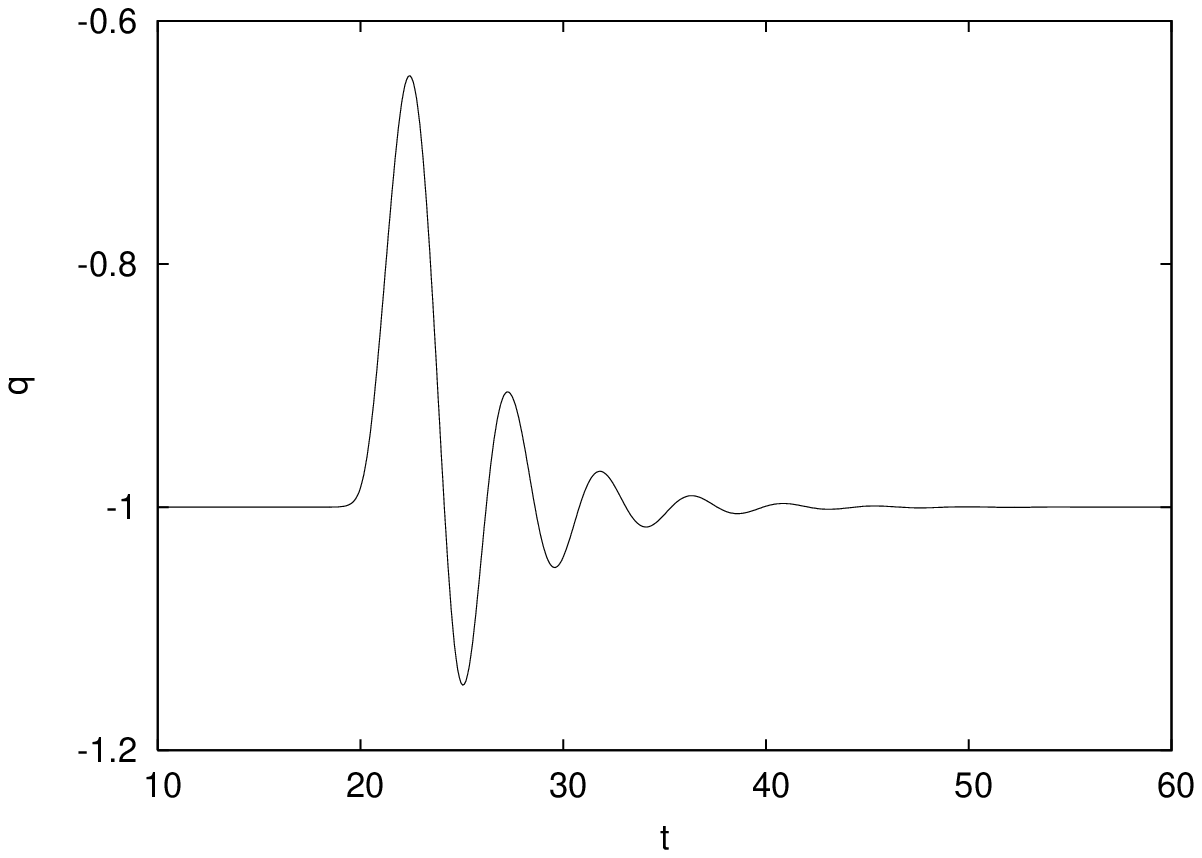,height=6cm,width=6cm,angle=0}
\psfig{figure=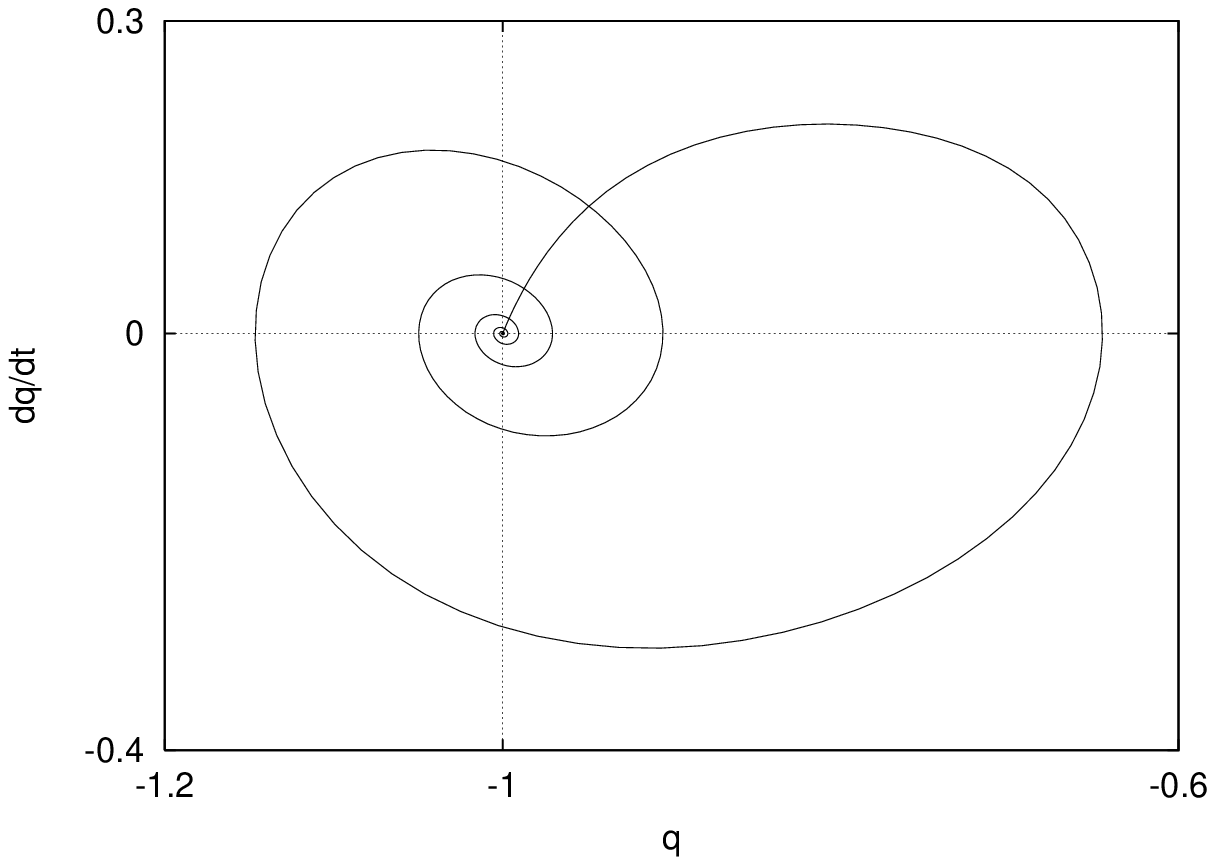,height=6cm,width=6cm,angle=0} }
\caption{Plot of $q(t)$ (left panel) and phase portrait $(q,dq/dt)$
(right panel) for a incident pulse below threshold so that the medium does
not switch. The parameters are $A_0 = 0.2,~ t_1=-20,~ t_2 =-22,~t_f=0.5,~ 
\gamma=1 $.}
\label{fig2}
\end{figure}
\begin{figure}[tbp]
\centerline{\psfig{figure=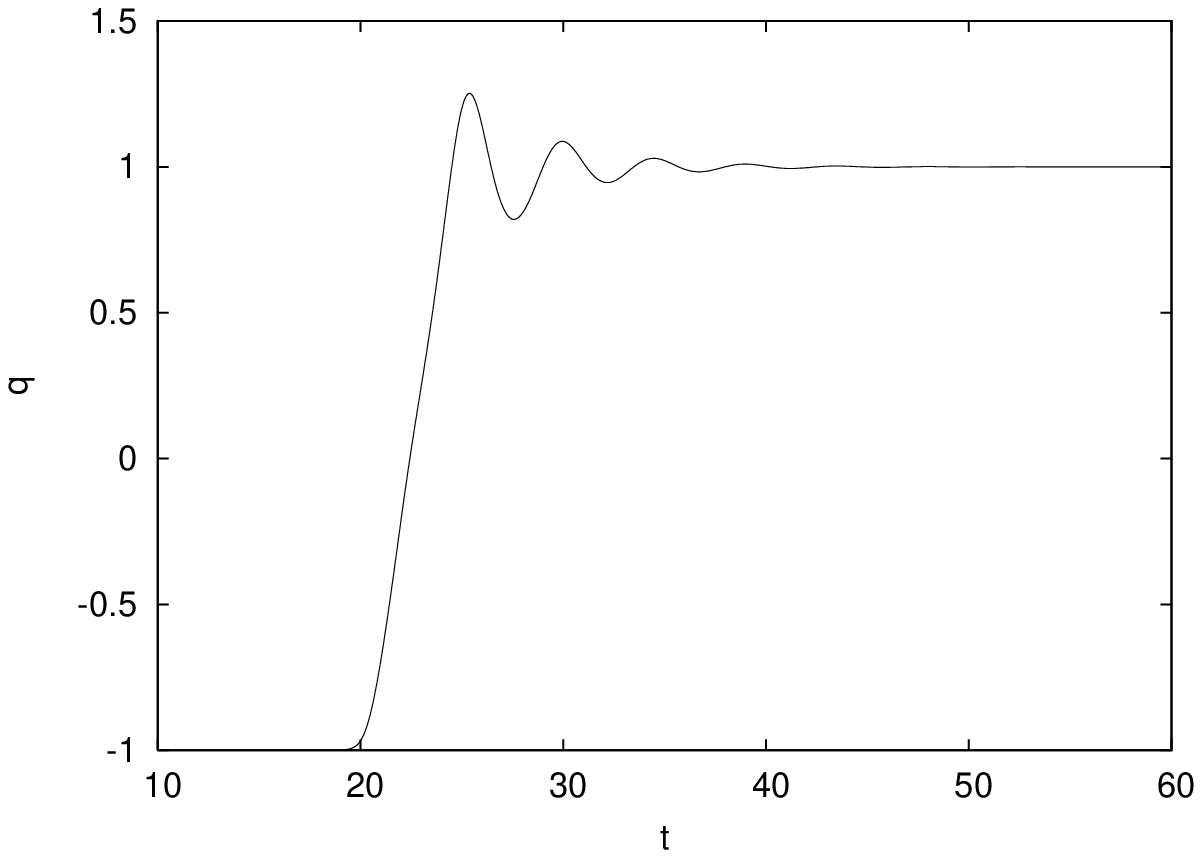,height=6cm,width=6cm,angle=0}
\psfig{figure=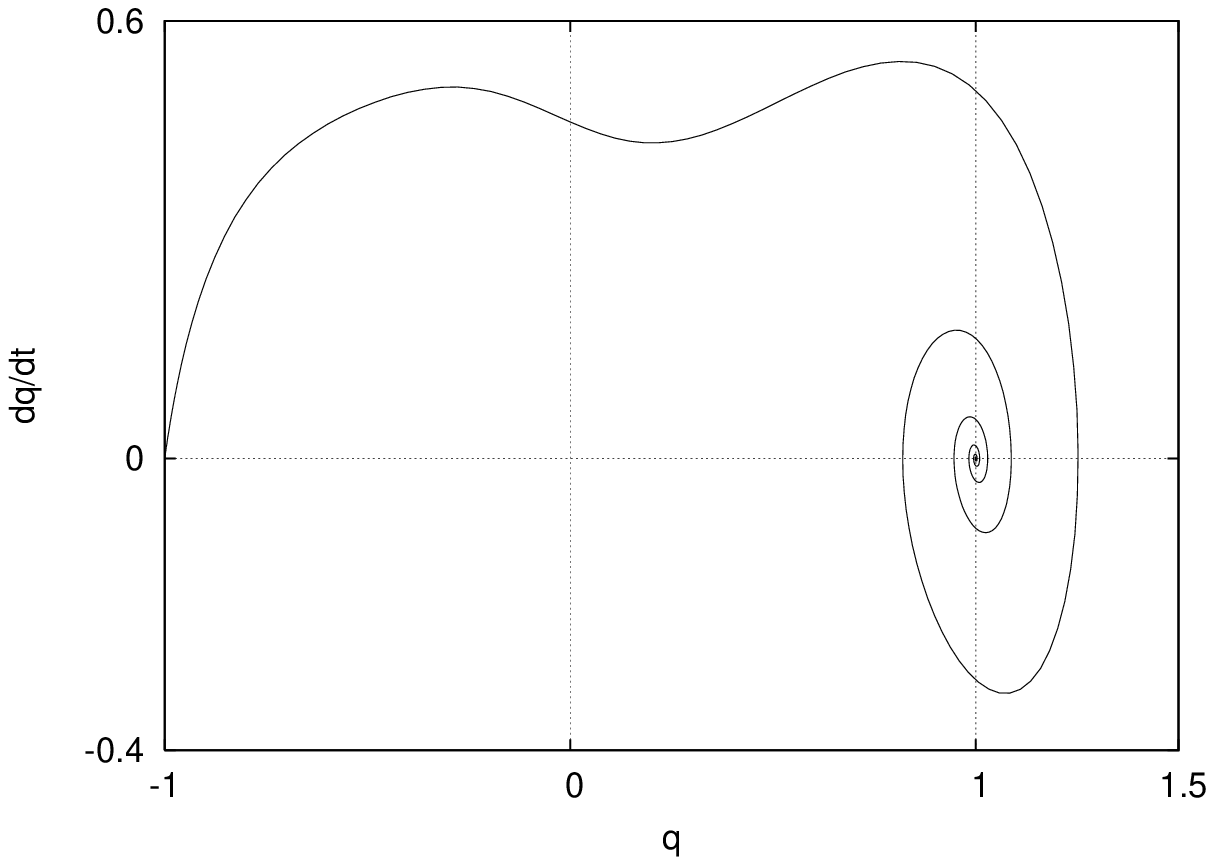,height=6cm,width=6cm,angle=0}}
\caption{ Same as fig. \protect\ref{fig2} but with a larger amplitude
$A_0 = 0.4$ so that switching occurs.}
\label{fig3}
\end{figure}

For a short electromagnetic pulse the
equilibrium positions can be considered as fixed. After the interaction, the
system evolves as if free.
Figure \ref{fig2} shows the evolution of the medium polarization and 
the corresponding phase plane under the action of an ultrashort 
electric field. The amplitude is not enough for switching and the system
relaxes to ${\bar q}=-1$ following the linearized behavior $\exp(i \omega t)$ where
$\omega$ is given by (\ref{poles2}).
If the amplitude is increased as in Fig. \ref{fig3} the system switches
to the fixed point ${\bar q}=1$ following again the linearized behavior (\ref{poles2}).

\subsection{Slowdown of switching}

\begin{figure}[tbp]
\centerline{\psfig{figure=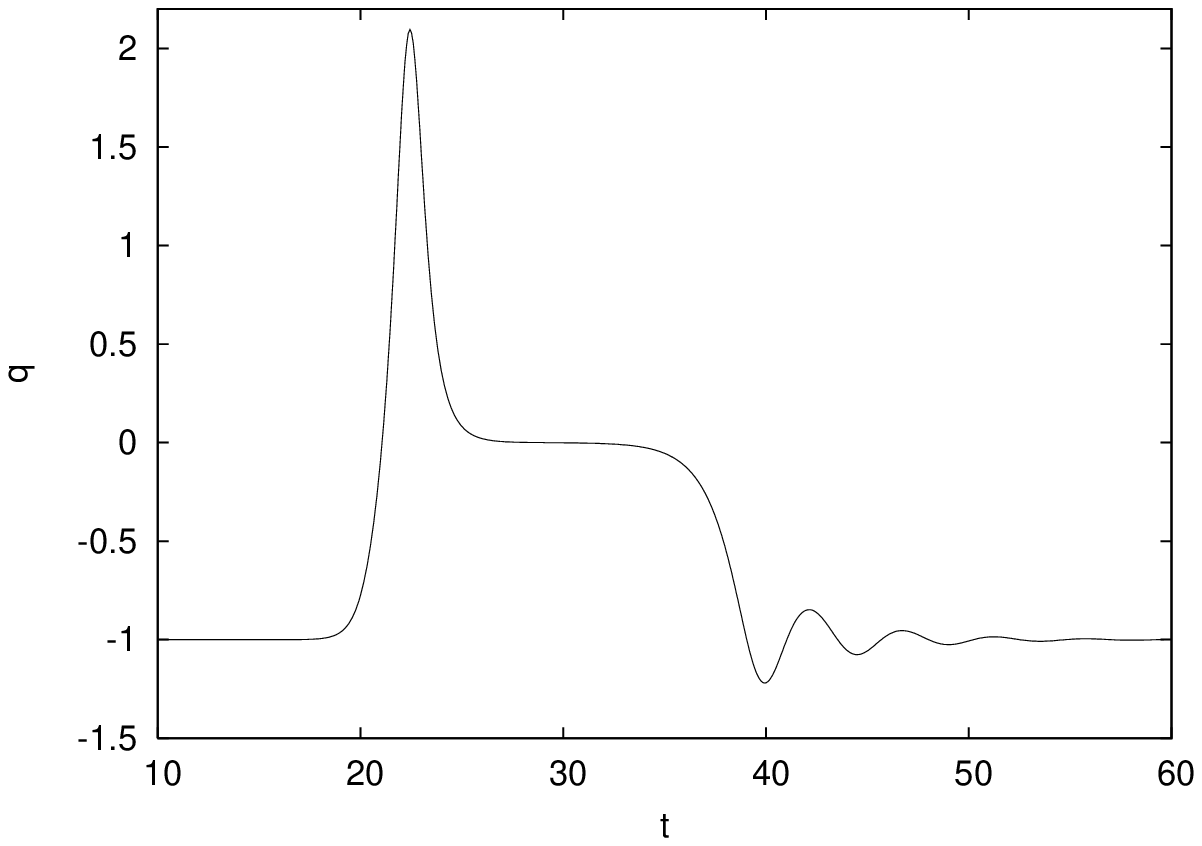,height=6cm,width=6cm,angle=0}
\psfig{figure=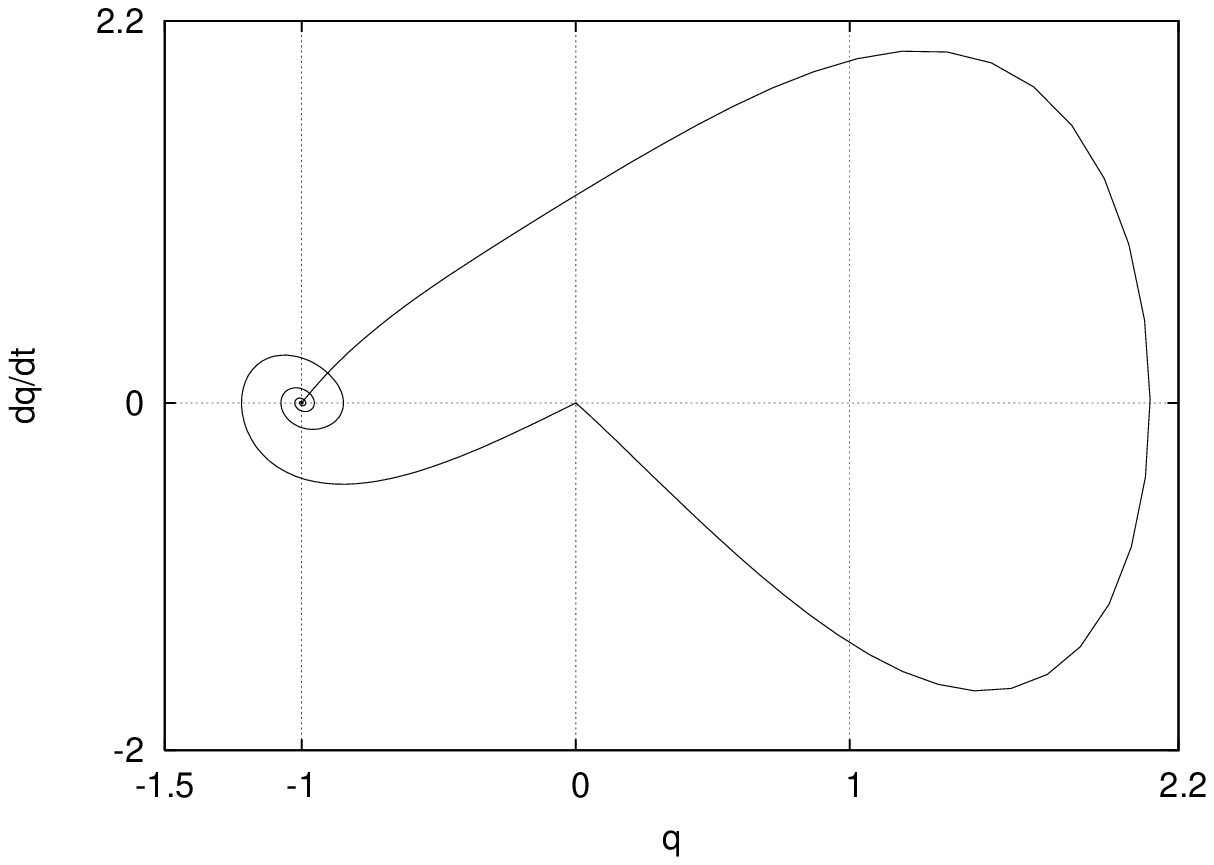,height=6cm,width=6cm,angle=0} }
\caption{Plot of $q(t)$ (left panel) and phase portrait $(q,dq/dt)$
(right panel) for a incident pulse below threshold so that the medium does
not switch. The parameters are $A_0 = 1,~ t_1=-20,~ t_2 =-22.988 ,~ \protect%
\gamma=1 $.}
\label{fig4}
\end{figure}

\begin{figure}[tbp]
\centerline{\psfig{figure=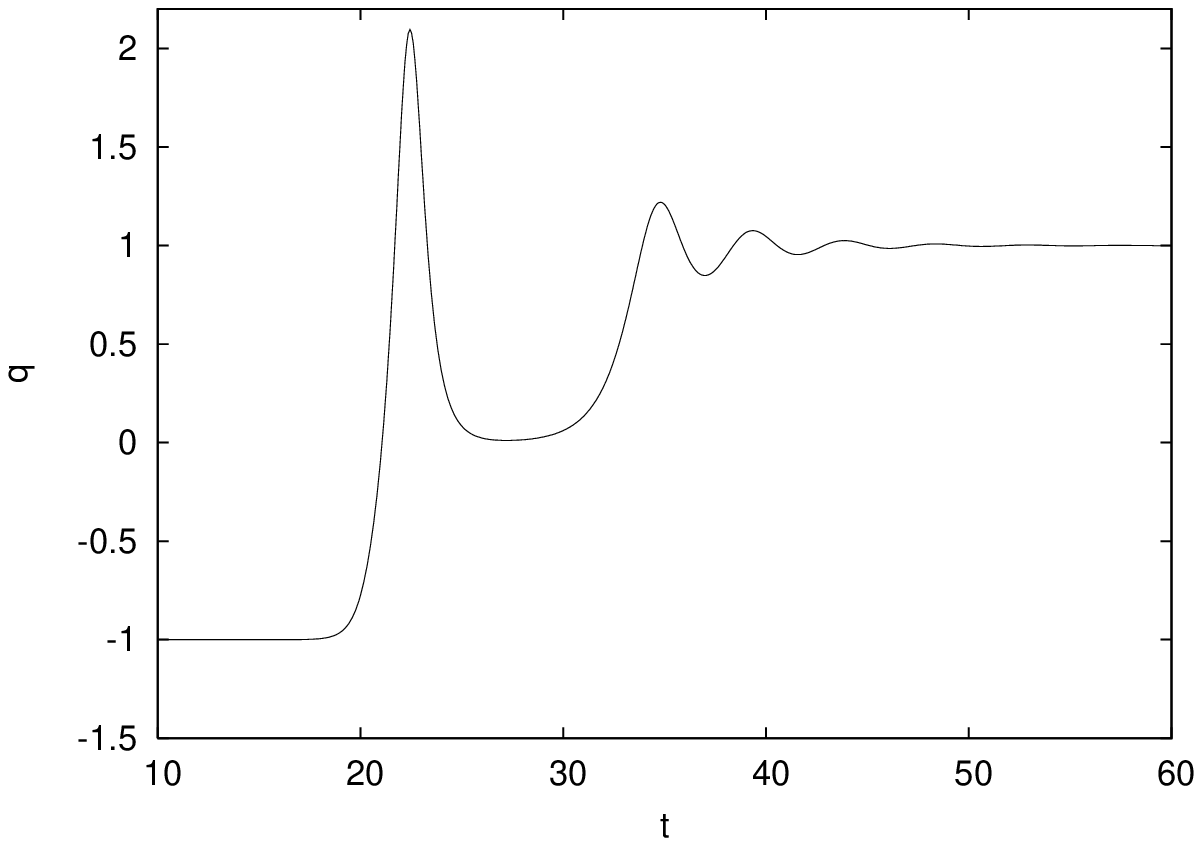,height=6cm,width=6cm,angle=0}
\psfig{figure=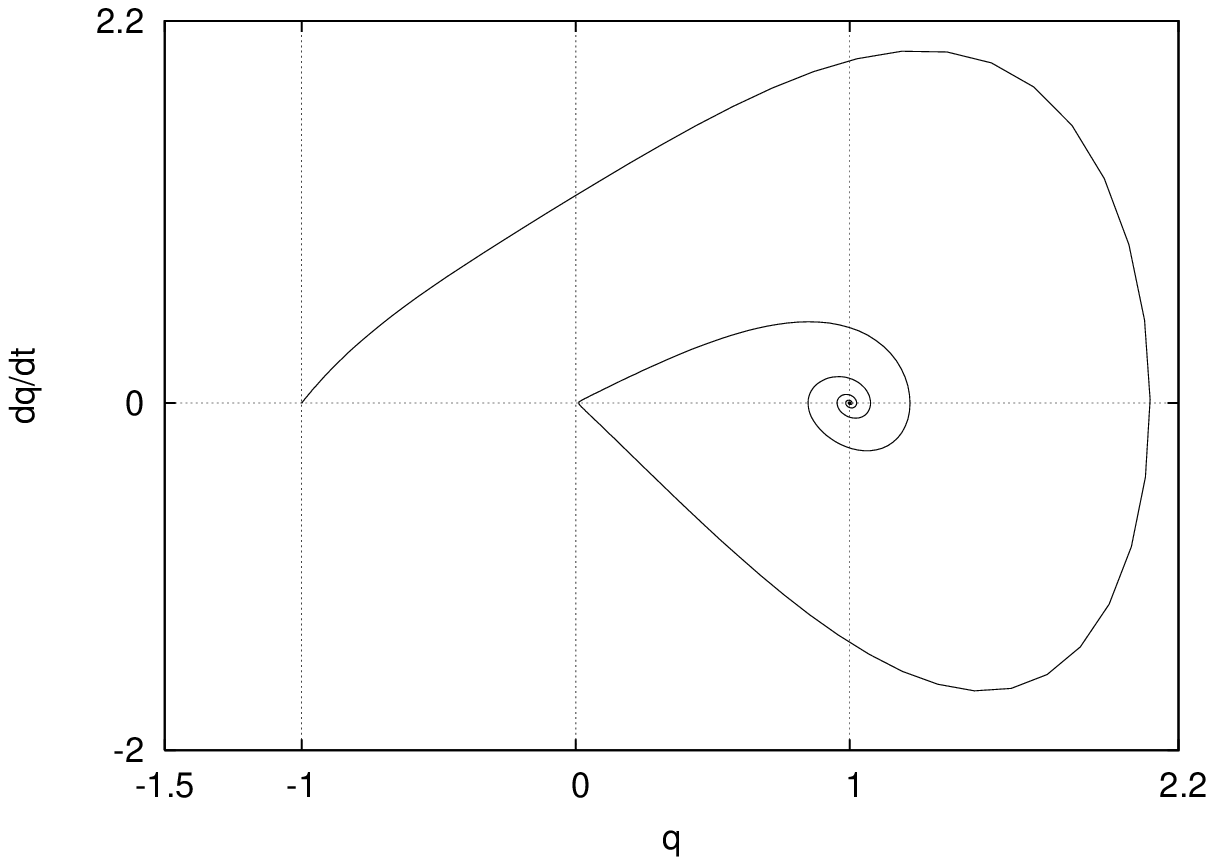,height=6cm,width=6cm,angle=0}}
\caption{ Same as fig. \protect\ref{fig4} but with a longer pulse $t_2
=-22.989$ so that switching occurs.}
\label{fig5}
\end{figure}

Now we consider the special cases where the incoming pulse brings the
system on a trajectory that goes near the unstable fixed 
point ${\bar q}=0$. Fig. \ref{fig4}
shows such a case below threshold. The left panel of Fig. \ref{fig4}
shows the slowing down of $q(t)$ around $q=0$ because there $dq/dt\approx 0$. 
As seen from the plots, the system remains "frozen" near the unstable
equilibrium state $q=0$ for a certain time $T_{del}$. In Fig. \ref{fig5}
we present the results for a slightly duration where the system has switched.

\begin{figure}[tbp]
\centerline{
\psfig{figure=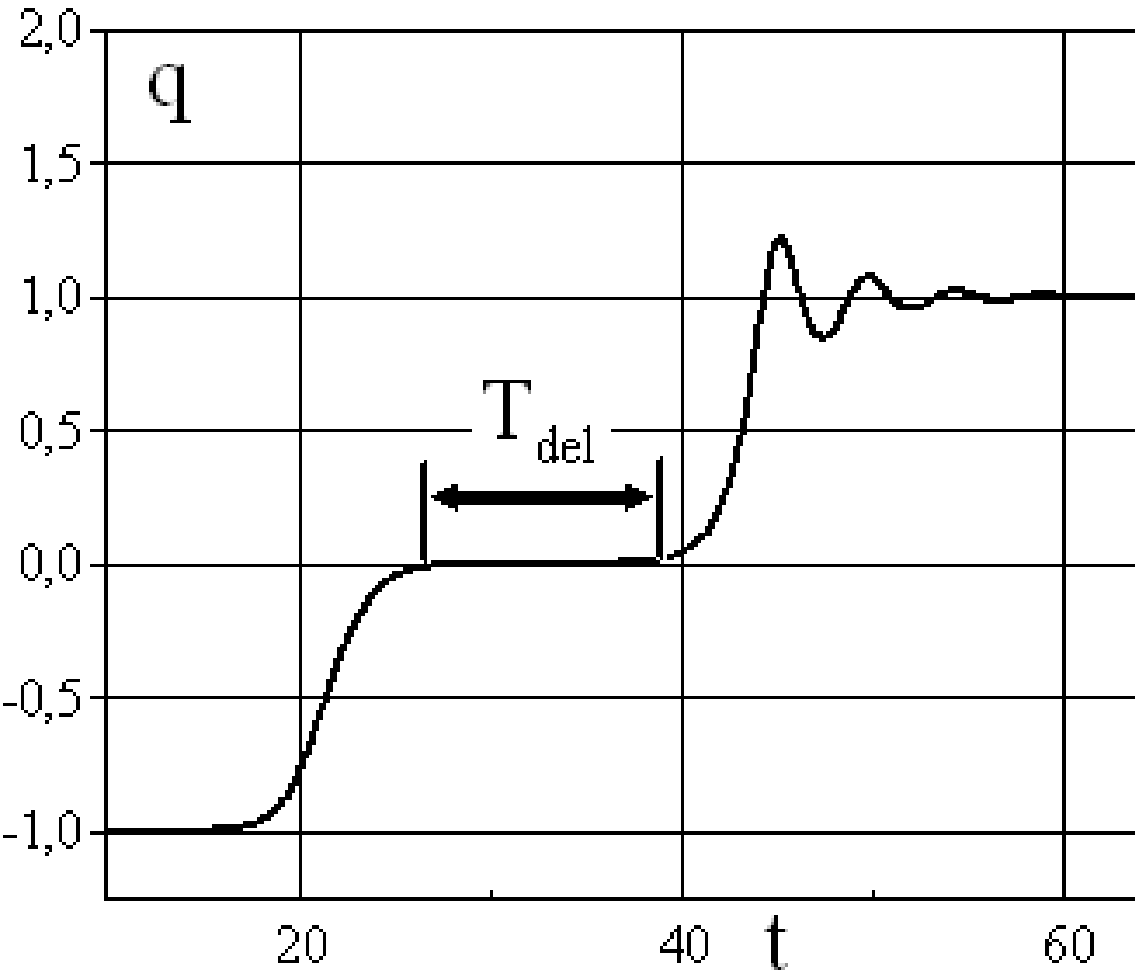,height=6cm,width=5cm,angle=0}
\psfig{figure=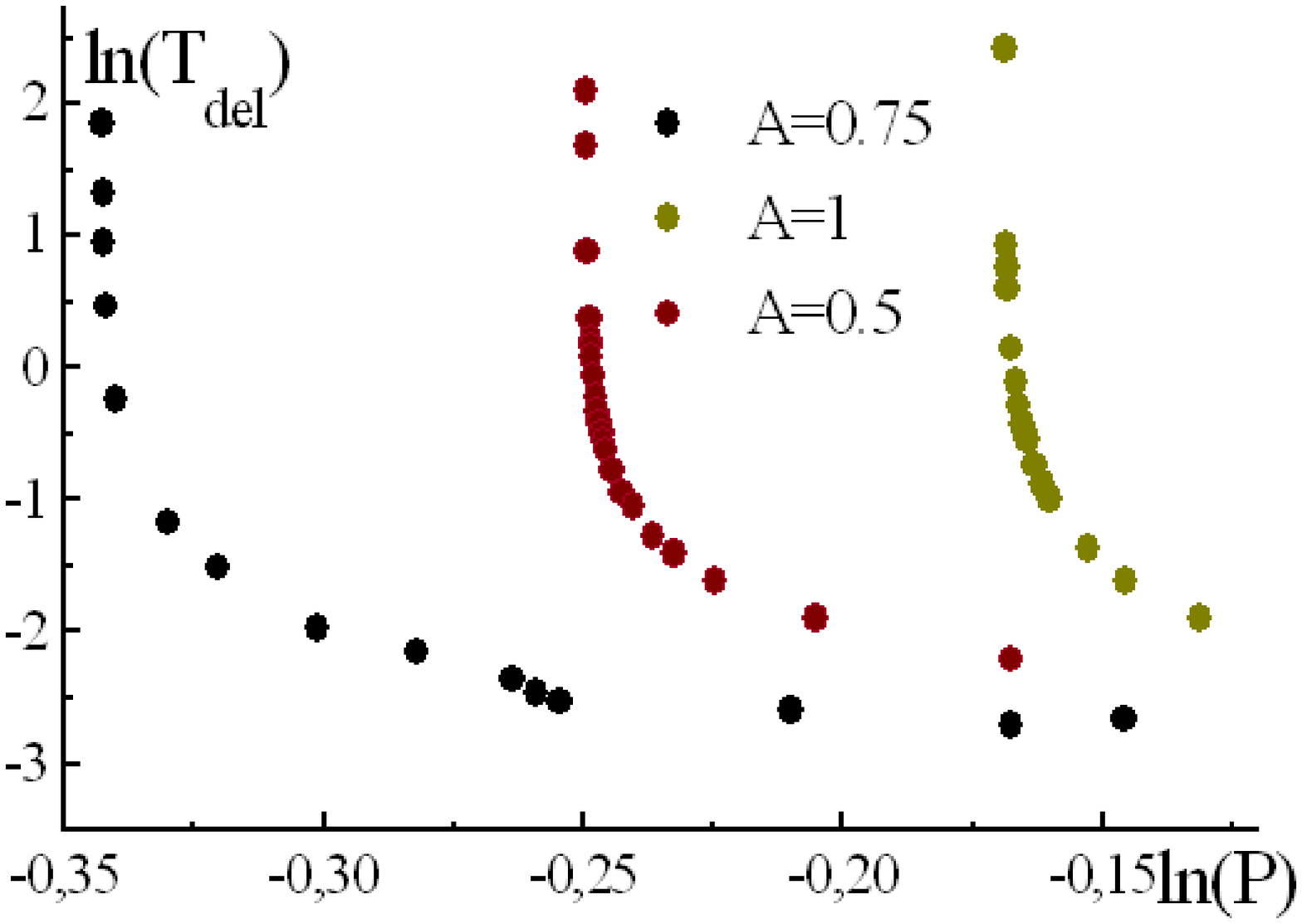,height=6cm,width=5cm,angle=0}}
\caption{Plot of $q(t)$ (left panel) showing the slowing down as the system
approaches the unstable fixed point $\bar q=0$. The right panel shows the
duration of the plateau $T_{\mathrm{del}}$ vs the power $P$ of the incident
pulse in log-log scale}
\label{fig6}
\end{figure}

To investigate this slowing down, we introduced a gaussian initial
electric field
\[
e_{0}(-t)=A_0\exp [-t^{2}/a^2]  .
\]
In Fig. \ref{fig6} we plot a typical evolution $q(t)$ indicating $T_{del}$
on the left panel. On the right panel we give $\log(T_{del}$ vs $\log(P)$
where $P$ is the total power in the electromagnetic pulse
$$P \equiv \int_{-\infty}^{+\infty}dt e^2(t) = A_0^2 a \sqrt{\pi\over 2}.$$
We present three different amplitudes $A_0=1,~0.75$ and $0.5$ and vary $P$
by varying the width $a$. Fig. \ref{fig6} shows that there are critical
values of power where the time delay becomes very large. There does not
seem to be a single critical index for the description of the singular behavior
of the time delay.

\subsection{Free evolution of the oscillator}

\begin{figure}[tbp]
\centerline{
\psfig{figure=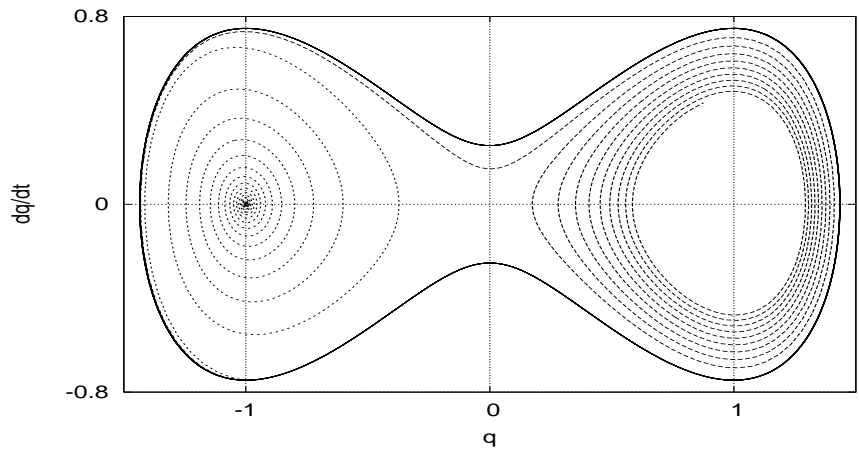,height=6cm,width=12cm,angle=0}}
\caption{Phase portrait $(q,{\dot q})$ showing free evolution of the
system for a delta function like electromagnetic pulse so that
$q(t=0) = -1,~~  dq/dt(t=0)=-0.75$. The long
dash corresponds to $\gamma=0.2$ and the short dash to $\gamma=0.5$.
The continuous curve corresponds to $\gamma=0$.}
\label{fig7}
\end{figure}

We consider the limit when the external electromagnetic field is so
short that it just gives an impulse to the oscillator which then
evolves as free. Then $e_0(t) = A_0 \delta(t)$.
Plugging this into (\ref{ferroq}), integrating in a small
interval around $t=0$ and assuming
continuity of $q$ we obtain $q_t|_{t=0^+}= \gamma A_0$.
The free evolution of the oscillator is then
governed by the equations%
\begin{eqnarray*}
dq/dt &=&p, \\
dp/dt &=&q- q^{3}-0.5\gamma ^{2}p,
\end{eqnarray*}
with the initial conditions $q(t=0)=-1,~~q_t(t=0)=\gamma A_0$.
The phase trajectories are solution of the equation%
\[
p\left( dp/dq\right) =q- q^{3}-0.5\gamma ^{2}p.
\]
The phase planes corresponding to this free evolution of the Duffing oscillator 
for $\gamma=0.5$ and $0.2$ are presented in Fig. \ref{fig7} together with the
nondamped situation. For $\gamma=0.5$ the damping is strong and prevents 
switching while for $\gamma=0.2$ the system can escape to the other equilibrium
and slowly converge to it.
The pictures on Fig. \ref{fig7} show that dissipation leads to damping of oscillation
around equilibrium position and supports the switching from one to another
equilibrium state. 
This dissipation results from the 
radiation of the electromagnetic waves out of the thin film.

\subsection{Long electromagnetic pulse}

When the electromagnetic pulse has the form of a plateau with a sharp
front and a sharp tail, transient steady states of the ferroelectric 
are created.
We can find them by considering 
the static solutions of equation (\ref{ferroq}):%
\begin{equation}
q- q^{3}+\gamma e_{0}=0.  \label{steady}
\end{equation}
Let $e_{0}$ be positive. If $0\leq \gamma e_{0}<2/(3\sqrt{3})$ there
are three roots, corresponding each to a fixed point, two stable
and one unstable. These can be calculated by perturbation 
when $\gamma e_{0}<<2/(3\sqrt{3})$, we have 
\be\label{fix3} \bar{q}_{0}(e_{0})\approx -\gamma e_{0},~~ 
\bar{q}_{1,2}(e_{0}) \approx \pm 1+\gamma e_{0}/2.  \ee
When $\gamma e_{0}=2/(3\sqrt{3})$ the unstable and left stable equilibrium
points merge together and we have only one stable point
\be\label{fix2} \bar{q}_{2}(e_{0})=2/\sqrt{3}.  \ee
In the limit of very big amplitude of electromagnetic pulse, when
$\gamma e_{0}>>2/(3\sqrt{3})$, there is only one fixed point which is
stable and is defined by the approximate formula
\be\label{fix1} \bar{q}_{2}(e_{0})\approx \left( \gamma e_{0}\right) ^{1/3}\left[ 1+%
\frac{1}{2}\left( \gamma e_{0}\right)
^{-2/3}\right] .  \ee

The numerical simulation of switching under the influence of long 
plateau-like pulse
demonstrates the damping nutations near the stable points 
$\bar{q}_{1,2}(e_{0})$.
The kinetics of switching can be considered by using the linearized
equation (\ref{ferroq}) near the stable fixed points:%
\[
\delta q_{tt}+\frac{\gamma ^{2}}{2}\delta q_{t}+[3\bar{q}%
_{1,2}(e_{0})-1]\delta q=0,
\]
where $\delta q=q-\bar{q}_{1,2}(e_{0})$. The associated
characteristic equation give the decrement and frequency of 
nutations 
\be\label{freqtrans}
\Gamma _{0}=\gamma ^{2}/4,~~ 
\Omega _{0}=\sqrt{3\bar{q}_{1,2}^2(e_{0})-1-\gamma^{4}/16}.  \ee
The decrement 
is independent of the initial pulse amplitude while 
the frequency of nutation depends on it.

\begin{figure}[tbp]
\centerline{
\psfig{figure=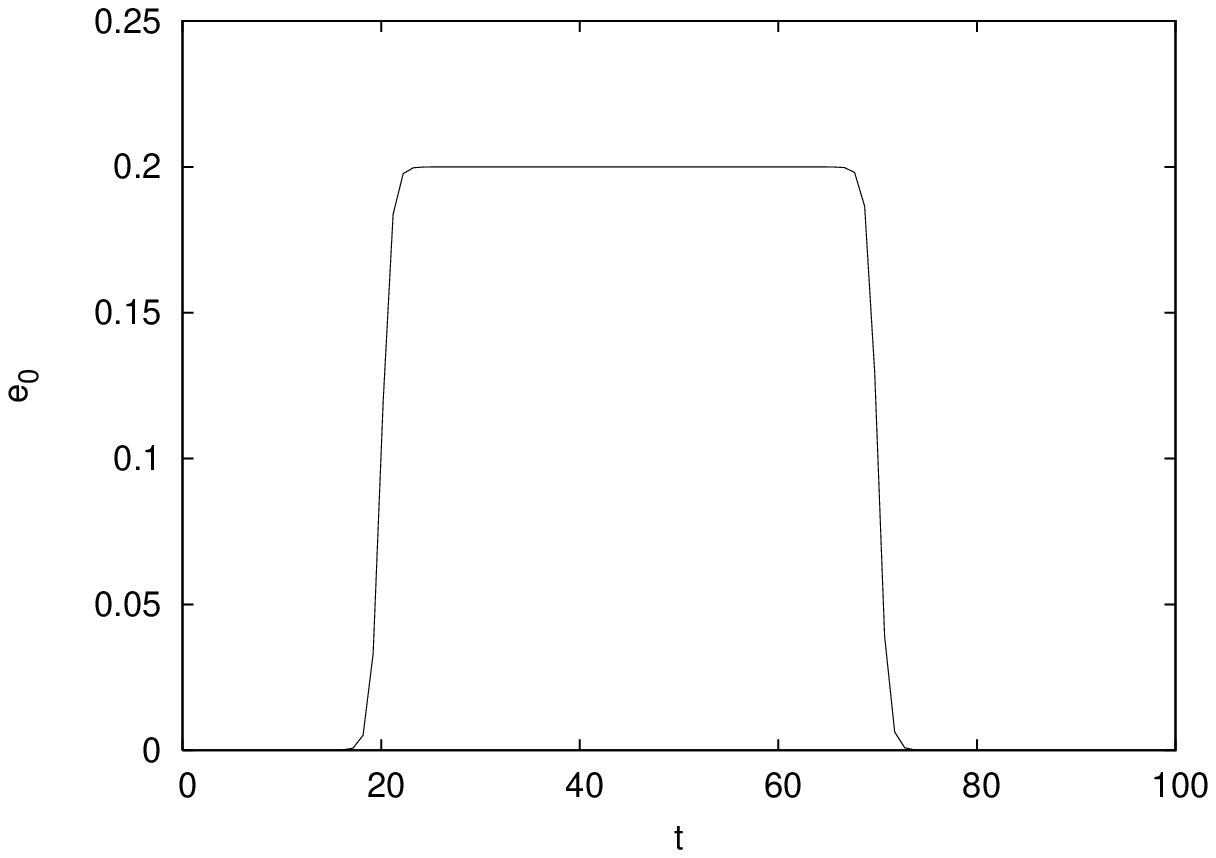,height=6cm,width=4cm,angle=0}
\psfig{figure=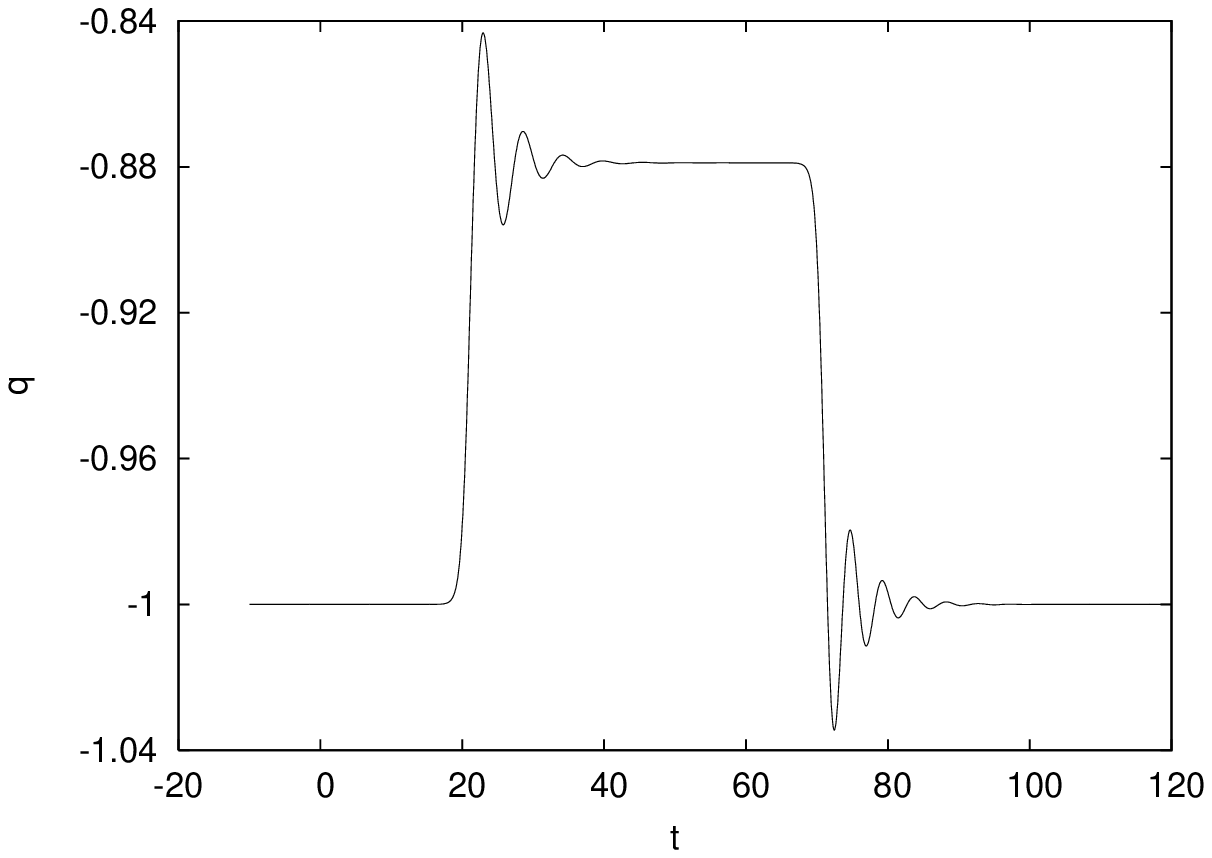,height=6cm,width=4cm,angle=0}
\psfig{figure=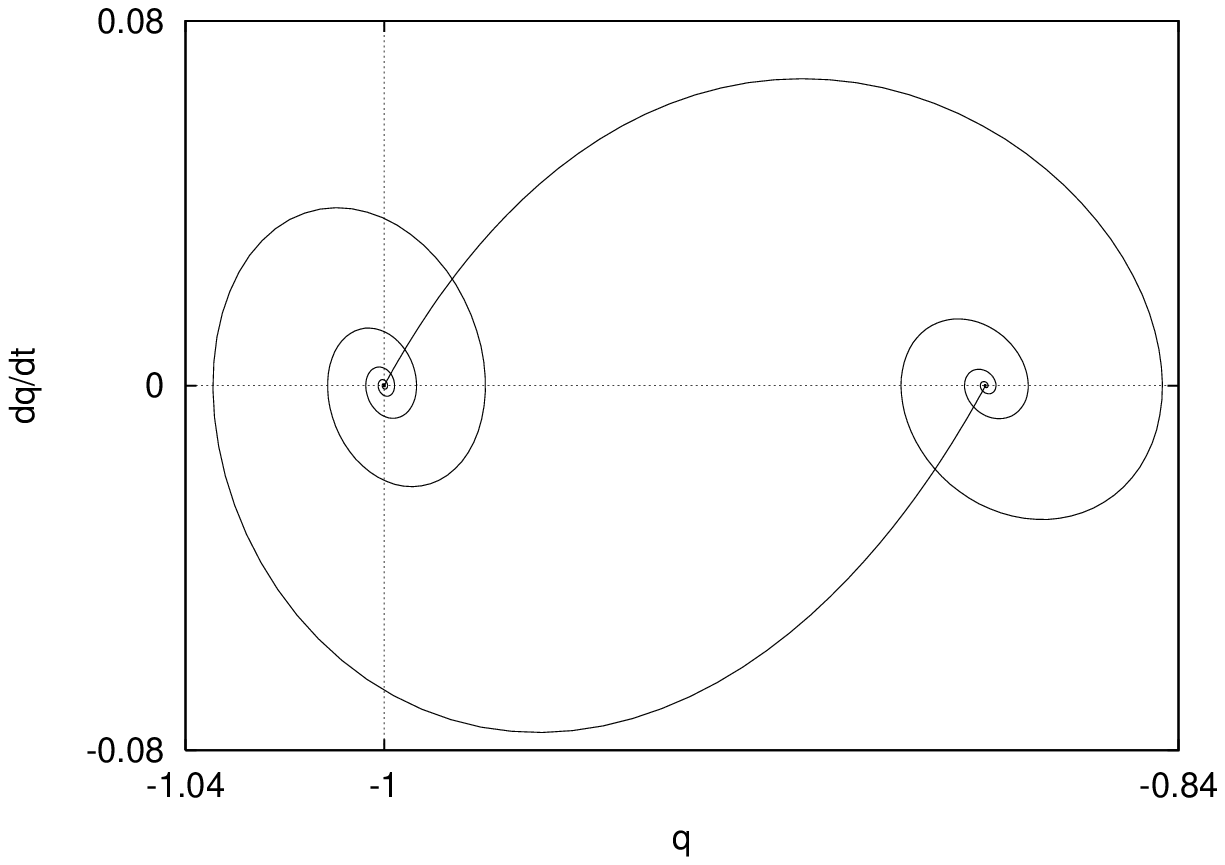,height=6cm,width=4cm,angle=0}}
\caption{Plot of a long incident pulse $e_{0}(x)$ (left panel), $q(t)$
(middle panel) and phase portrait $(q,dq/dt)$ (right panel). The
parameters are the same as in Fig. \protect\ref{fig4} except $%
A_{0}=0.1,~t_{1}=-70,~t_{2}=-20$. }
\label{fig8}
\end{figure}

\begin{figure}[tbp]
\centerline{
\psfig{figure=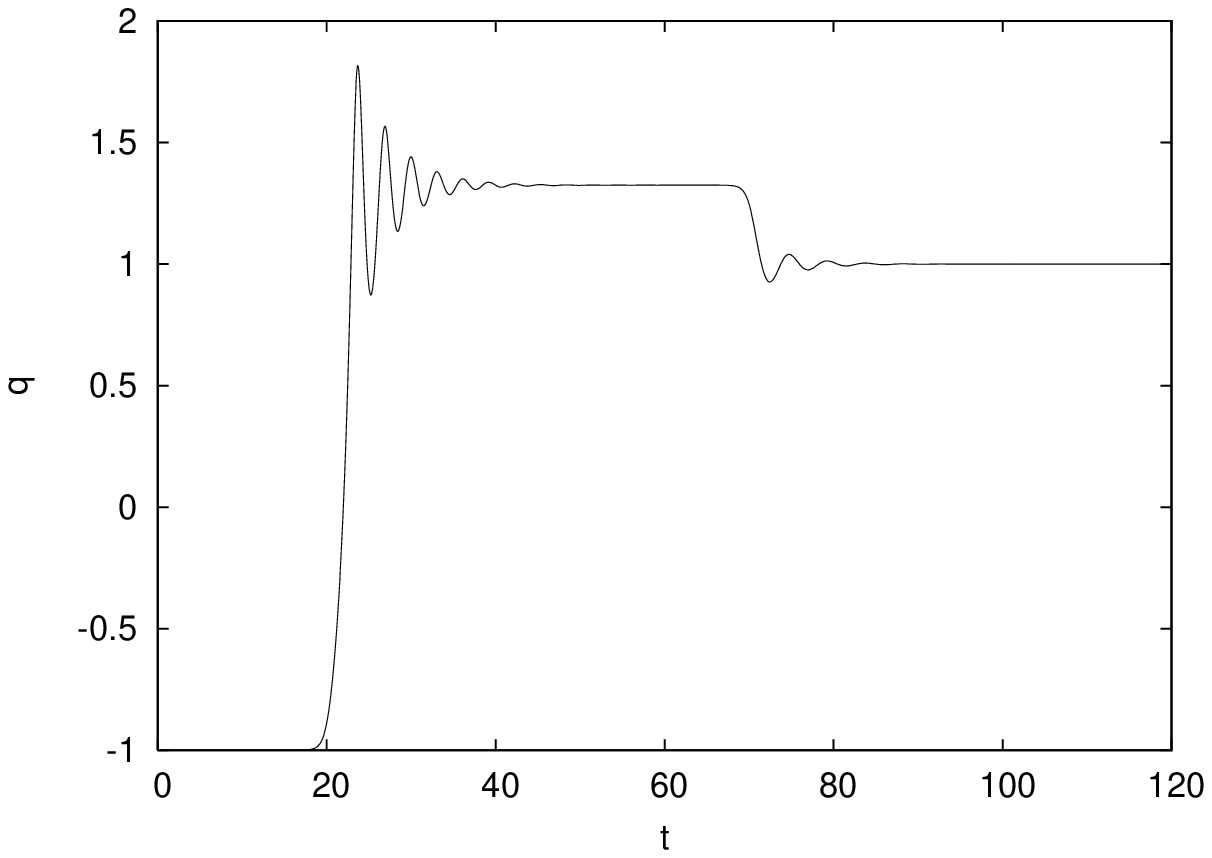,height=6cm,width=5cm,angle=0}
\psfig{figure=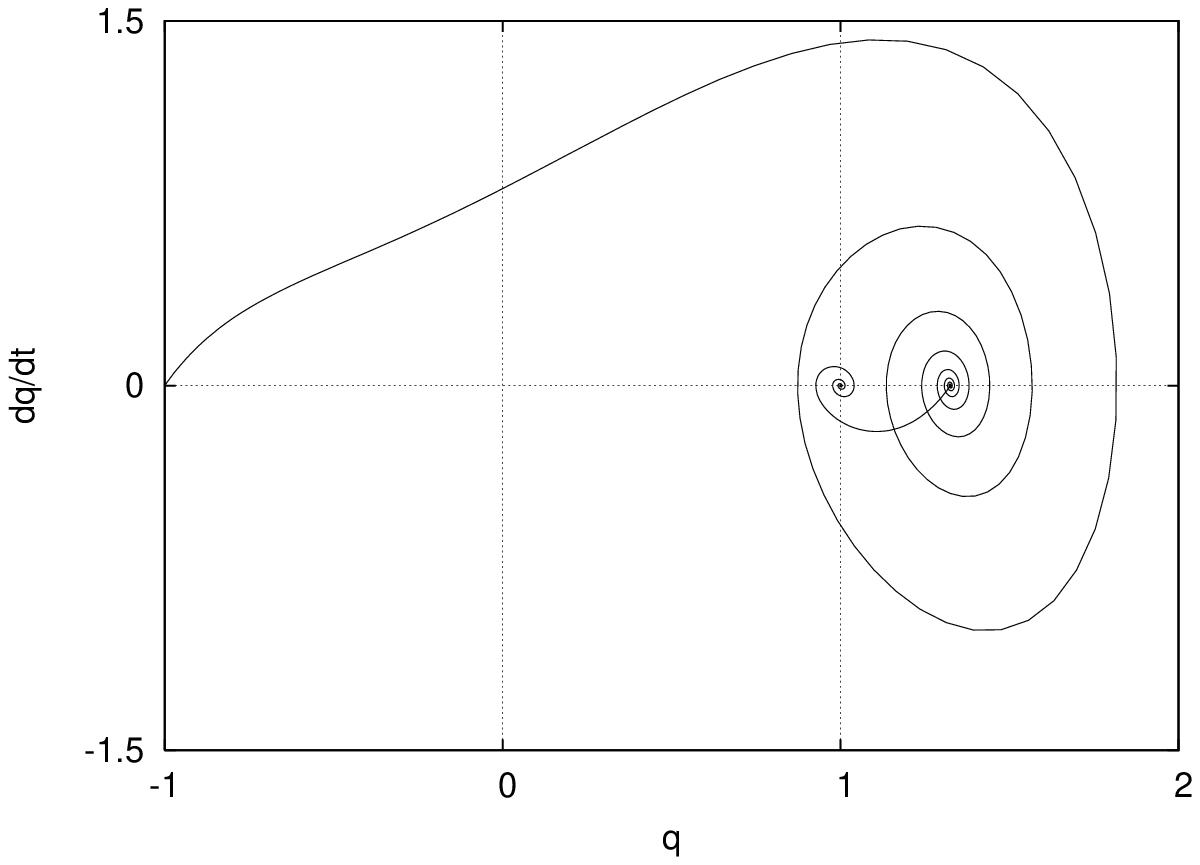,height=6cm,width=5cm,angle=0}}
\caption{Plot of $q(t)$ (left panel) and phase portrait $(q,dq/dt)$ for an
incoming long pulse similar to the one in Fig. \protect\ref{fig4} except
that $A_0=0.5$}
\label{fig9}
\end{figure}

The following evolution of the ferroelectric polarization depends on the
area of the external field. In this section we consider long pulses
that create transient equilibria in the ferroelectric.
Fig. \ref{fig8} shows in the left panel such a long pulse, in the
middle panel the response $q(t)$ of the medium and in the right panel
the associated phase plane. The ferroelectric is moved into a
transient equilibrium and returns
to its previous stable polarization state $\bar{q}=-1$. Here we chose
$A_0=0.1$ so that $\gamma e_0 = 0.2$ in the plateau region. The
system goes to the transient steady state ${\bar q_1}\approx -0.88 $
value that is in good agreement with (\ref{fix3}). which gives $-0.9$.
The decrement $\Gamma _{0}\approx 0.25$ and nutation frequency 
$\Omega _{0}\approx 2$ is
of the motion to this transient 
fixed point is predicted correctly by the estimates (\ref{freqtrans}).
When the system returns to its natural fixed point, the estimates are
again correct and given by (\ref{poles2}). Notice that the nutation 
frequency around the
transient fixed point is twice as big as the one around the 
natural fixed point.

In Fig. \ref{fig9}
we observe the same phenomenon except that the transient equilibrium
is close to the other stable polarization state $\bar{q}=+1$ so that
when the field returns to zero, the system relaxes to that state.
Here $A_0=0.5, \gamma =1$ so that in the plateau region $e_0=1$. 
We are in the region above the critical $e_0$ so that there is
only one fixed point.
The estimate (\ref{fix1}) gives ${\bar q}\approx 1.5$ which is in 
excellent agreement with the numerical value 1.4.

\section{Concluding remarks}

\begin{figure}[tbp]
\centerline{\psfig{figure=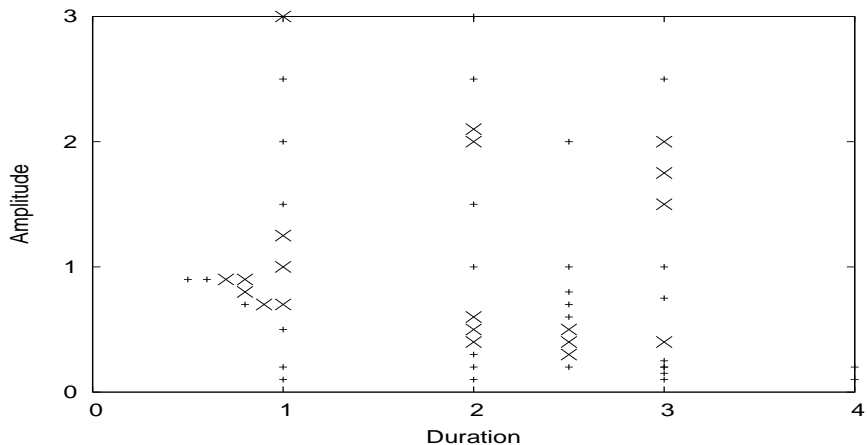,height=6cm,width=12cm,angle=0}}
\caption{Parameter plane (amplitude,duration) of incident pulse detailing
the final state of the film switched or non switched. The $+$ symbols
correspond to no switching and the $\times$ to switching.}
\label{fig10}
\end{figure}

We considered the simplest model of interaction of a short electromagnetic
pulse with a thin ferroelectric medium where the polarisation can be considered
as uniform. The duration of the electromagnetic pulse is much shorter
than the relaxation time of the medium so that only radiative decay occurs.

The linear scattering formalism predicts that low amplitude pulses can
be completely reflected by the medium, in a frequency band that grows
with the coupling $\gamma$. On the contrary strong electromagnetic fields
can switch the medium from one state to another. We have studied such
switching phenomena for both short and long pulses. For the latter we 
characterized the transient states they create.

We define the switching time $t_s$ as the time interval for the
system to go from one fixed point to the other. From the Duffing
system the typical (normalized) damping 
time is $\gamma^2/4$. If
$\gamma^2 > 4 $ switching occurs during the front of the pulse, otherwise
the switching time is of the order of $ \gamma^2/4$ because the system circles
around the fixed point before reaching it. In any case the system will
always switch in a time smaller or of the order of $\gamma^2/4$. 
In physical units this is about 
$$t_s = 4 |\alpha_0|^{-3/2} |T_c-T|^{-3/2}.$$
In some
cases, the switching time can be longer, in particular if the field
drives the ferroelectric near the unstable fixed point causing
considerable slowing down. Switching is also irregular as shown 
by Fig. \ref{fig10} which gives the events in the plane (duration,amplitude)
of the incoming pulse. There one sees that a threshold amplitude is
needed for switching. Above that the system switches or not depending on
the pulse duration. For a given duration there seems to be windows where
the system switches. 
All this information can be used by experimentalists to
estimate parameters.
Finally we believe the model due to its simplicity and generality can be 
transposed to other electromagnetic systems.

\section*{Acknowledgment}

It is a pleasure for the authors to thank Prof. S.O. Elyutin for very useful
discussions. A.I. Maimistov and E.V.Kazantseva are grateful to the \textit{%
Laboratoire de Math\'{e}matiques, INSA de Rouen} for hospitality and
support. This research was partially supported by RFBR under grant No:
06-02-16406.

\end{document}